\documentclass[twoside]{article}
\usepackage{fleqn,espcrc2,multicol}

\usepackage{graphicx}

\def\btg{\mbox{$\beta_T \gamma_T$}}

\def\ui{u_{Ti}}
\def\utm{\langle u_T \rangle}
\def\d{{\rm d}}
\def\vl{\Big\vert}
\def\bb{\mbox{\boldmath${\scriptscriptstyle\beta}$}}
\def\bbl{\mbox{\boldmath$\beta$}}
\def\ee{e$^+$e$^-\;$}
\def\ss{${\rm s}\bar{\rm s}\;$}
\def\ssb{$\langle {\rm s}\bar{\rm s}\rangle\;$}

\newcommand{\AmS}{{\protect\the\textfont2
  A\kern-.1667em\lower.5ex\hbox{M}\kern-.125emS}}

\title{
\vspace*{-4cm} 
\begin{small}
\begin{flushright}
{\underline{DFF 368/10/2000}} \\
\end{flushright}
\end{small}
\vspace*{2.6cm}
Transverse momentum spectra of identified particles
in high energy collisions with statistical hadronisation model}

\author{\underline{F. Becattini}, L. Bellucci and G. Passaleva
   \address{University of Florence and INFN Florence, 
    Largo E. Fermi 2, I-50125, Firenze, Italy}}
       
\begin{document}

\begin{abstract}
A detailed analysis is performed of transverse momentum spectra of 
several identified hadrons in high energy collisions within 
the framework of the statistical model of hadronisation. The effect 
of the decay chain following hadron generation is accurately taken into 
account. The considered centre-of-mass energies range from $\simeq$ 10 
to 30 GeV in hadronic collisions ($\pi$p, pp and Kp) and from $\simeq$ 
15 to 45 GeV in \ee collisions. 
A clear consistency is found between the temperature parameter extracted
from the present analysis and that obtained from fits to average hadron 
multiplicities in the same collision systems. This finding indicates 
that in the hadronisation, the production of different particle species
and their momentum spectra are two closely related phenomenons governed
by one parameter.
\vspace{1pc}
\end{abstract}

\maketitle

\section{Introduction}

The idea of a statistical approach to hadron production in high energy 
collisions dates back to '50s \cite{fermi} and '60s \cite{hagedorn} and 
it has been recently revived by the observation that hadron multiplicities 
in \ee collisions agree very well with a thermodynamical-like ansatz 
\cite{beca1,beca2}. 
This finding has also been confirmed in hadronic collisions and it has 
been interpreted in terms of a pure statistical filling of multi-hadronic 
phase space of assumed pre-hadronic clusters (or fireballs) formed in 
high-energy collisions, at a critical value of energy density \cite{beca3}. 
In this framework, temperature and other thermodynamical quantities 
have a statistical meaning which does not entail the existence of a 
thermalised hadronic system on an event-by-event basis.
Stated otherwise, statistical equilibrium shows up only when comparing many 
different events, whilst in each of them the Gibbs law of equally likely 
multi-hadronic states applies.
So far, this proposed statistical cluster hadronisation model has been 
mainly tested against measured abundances of different hadron species 
for a twofold reason. Firstly, unlike momentum spectra, they are quantities 
which are not affected by hard (perturbative) QCD dynamical effects but are 
only determined by the hadronisation process; indeed, in the framework of
a multi-cluster model, they are Lorentz-invariant quantities which are 
independent of the cluster's overall momentum. Secondly, they are quite easy
to calculate and provide a very sensitive test of the model yielding an 
accurate determination of the temperature. However, in order to establish 
the validity of this approach, it is necessary to test further observables 
and to assess their consistency with the results obtained for multiplicities. 
Certainly, one of the best suited observables is the transverse momentum of 
identified hadrons, where transverse is meant to be with respect to beam 
line in high energy hadronic collision, and thrust or event axis in high 
energy \ee collisions. Indeed, such projection of particle momentum is 
supposed to be the most sensitive to hadronisation or, conversely, the 
least sensitive to perturbative QCD dynamics.\\ 
Actually, it has been known for a long time that transverse momentum 
distributions are Boltzmann-like in hadronic collisions and this very 
observation was pointed out by Hagedorn as a major indication in favour 
of his statistical model of hadron production \cite{hagedorn2}. It must 
be emphasized that the prediction of a thermal-like 
shape in principle only applies to particles directly emitted from the 
hadronising source, whereas measured spectra also include particles produced 
by decays of heavier hadrons. However, most analyses do not take into account the 
distortion of primordial hadronisation spectrum due the to hadronic decay 
chain and try to fit the data straight through it. This problem has been 
discussed in literature \cite{hagedorn3} and an analytical calculation 
has been developed to take into account the effect of two and three body 
decays \cite{sollfran,sollfran2}, which has then been used both for 
pp \cite{sollfran2} and heavy ion collision 
\cite{sollfran,sollfran2,wiedheinz} including most abundant resonances. 
In this paper we introduce a method allowing to rigorously and exhaustively
determine the contribution of all particle decays. Hence, by taking 
advantage of this technique, we have performed an analysis of many measured 
transverse momentum spectra of identified hadrons in a wide range of 
centre-of-mass energies for several kind of collisions. 

\section{Statistical hadronisation and transverse momentum spectra}

The statistical hadronisation model \cite{beca4} assumes that in high energy 
collisions, as a consequence of strong interaction dynamics, a set of 
colourless clusters (or fireballs) is formed having certain values of mass, 
volume, internal quantum numbers and momentum, the latter being inherited 
from the hard stage of the process. Those clusters are assumed to give rise 
to hadrons according to a pure statistical law in the multi-hadronic phase 
space defined by their mass, volume and quantum numbers. This approach differs 
from another popular cluster hadronisation model \cite{webber} mainly because 
it gives clusters a volume so that hadron production is ruled by the properly 
understood phase space rather than relativistic momentum space. In this framework, 
the use of statistical mechanics and thermodynamical quantities, such as 
temperature, which need spacial dimension besides momentum space in order to 
be meaningful, is allowed.
We emphasize once more that the introduction of such quantities does not
entail any thermalisation process of hadrons after their formation, nor
the existence of a thermalised system event by event.\\ 
Although many clusters with different momenta, volumes, masses and quantum 
numbers are formed, it can be shown that the average values of many observables, 
e. g. particle multiplicites, are the same as those relevant to one 
equivalent cluster having suitable values of volume (namely the sum of all 
cluster volumes measured in the rest frame of the equivalent cluster itself), 
mass and quantum numbers (namely the sum of all cluster quantum numbers).
The proof of this statement \cite{prep}, a lengthy one, requires the assumption 
of special probabilities governing the fluctuations of cluster masses and quantum
numbers for a given set of volumes. If the volume (or mass) of the equivalent 
cluster is large enough, it is then reasonable to take a canonical approach (i.e. 
introducing a temperature) in order to calculate mean quantities, instead of 
carrying out an involved microcanonical calculation. Therefore, even though actual
clusters are small sized and microcanonical calculations would be needed
to determine mean quantities within each of them, the choice of suitable
mass fluctuation probabilities for the clusters allows one to calculate overall 
means dealing with only one large global structure and with much
fewer parameters. Arguing the other way around, it is not difficult to be 
convinced that the fluctuation probabilities to be chosen in order to achieve
such reduction of the problem are exactly the same as those of obtaining a 
given set of masses by splitting a large cluster into a given number of clusters
with volumes $(V_1,\ldots,V_N)$. This statement extends a reduction procedure 
to the microcanonical case which was proved in the canonical case \cite{beca3}, 
where, from the very beginning, clusters were given temperature and volume 
instead of mass and volume.  
Nevertheless, it is not obvious how large the equivalent cluster should be for 
the canonical approximation to apply. For the present, we have adopted a 
simple-minded {\it a posteriori} method consisting in justifying the canonical 
framework by its capability of accordance with the data. A simple argument to 
support the canonical approximation even at moderate values of volumes is the 
very large number of states ($O(10^2)$), i.e. hadrons and resonances, which 
can be excited in a hadron gas.\\
As far as single-particle transverse momentum spectra are concerned,
a similar reduction theorem from many clusters to one equivalent cluster 
in the averaging procedure applies \cite{prep}, though only approximately. 
In general, it can be shown that the {\em primary} spectrum of $j^{\rm th}$ 
hadron species depends on transverse four-velocities $\ui = \btg_i$
\footnote{In fact, we mean by transverse four-velocity the module of the
spacial part of a velocity four-vector $u=(\gamma, \bbl\gamma)$ with vanishing
longitudinal component.} of the $N$ clusters: 

\begin{eqnarray}\label{one}
  && \frac{\d n_j}{\d p_T} = \Big[ \prod_{i=1}^N \int_0^\infty \d \ui \Big] 
  \; f(u_{T1},\ldots,u_{TN}) \nonumber \\
  && \times \; \sum_{i=1}^N \frac{\d n_j}{\d p_T}\vl_{i}(\ui) \, .
\end{eqnarray}
where $f(u_{T1},\ldots,u_{TN})$ is the transverse four-velocities
distribution function. If one expands all single-cluster spectra 
$\d n_j/\d p_T|_{i}(\ui)$ in series of $\ui$ starting from a common 
value $\utm$ for all clusters, it can be proved that, at the zeroth order, 
the spectrum in eq.~(\ref{one}) becomes the same as that obtained for the 
aforementioned equivalent cluster, endowed with a tranverse four-velocity 
$\utm$. This reduction possibly allows taking the canonical approach in 
order to calculate the transverse momentum spectrum since the equivalent 
cluster size is much larger than single cluster's:   

\begin{eqnarray}\label{two}
  && \frac{\d n_j}{\d p_T} \propto \frac{(2J_j+1)}{\sqrt{1+\utm^2}} \,
   m_T p_T \nonumber \\
  && \times \; {\rm K}_1 \Big( \frac{\sqrt{1+\utm^2}\, m_T}{T} \Big) 
    \; {\rm I}_0 \Big( \frac{\utm p_T}{T} \Big)  
\end{eqnarray} 
where $m_T = \sqrt{p_T^2+m_j^2}$, $T$ is the temperature, and ${\rm K}_1$,
${\rm I}_0$ are modified Bessel functions. Eq.~(\ref{two}) is the 
Boltzmann limit of quantum statistics and it is a very good approximation
for all hadrons except pions \cite{beca1}.
For the zeroth order approximation in the $\ui$ expansion to be sufficiently 
accurate, the involved transverse four-velocities should be small, i.e. 
$f(u_{T1},\ldots,u_{TN}) \rightarrow 0$ already for $\ui \ll 1$. Since 
it is possible to choose the starting point $\utm$ of the series expansion
to make the first order term vanishing \cite{prep}, the largest neglected term
turns out to be $O((\ui-\utm)^2)$ which is very small provided that the 
aforementioned condition on $f$ is met. The analysis presented in this paper 
assumes the validity of this approximation and primary spectra have been 
calculated according to eq.~(\ref{two}).\\ 
Even though eq.~(\ref{two}) looks like the spectrum from a thermalised source
with a superimposed flow, a popular formula in the heavy-ion community, it is 
well worth emphasizing that in fact this formula has nothing to do with flow. 
Indeed, the average transverse four-velocity $\utm$ is not meant to be 
the mean value of cluster velocity distribution in a single event; 
rather, it is the average transverse four-velocity of all clusters over all 
collision events. In other words, $\utm$ may well arise from single clusters 
emitted at high $p_T$ in some events, e.g. following a hard parton scattering, 
and does not imply by any means a collective event-by-event expansion.
\begin{figure}[htb]
\vspace{5pt}
\includegraphics[width=17pc]{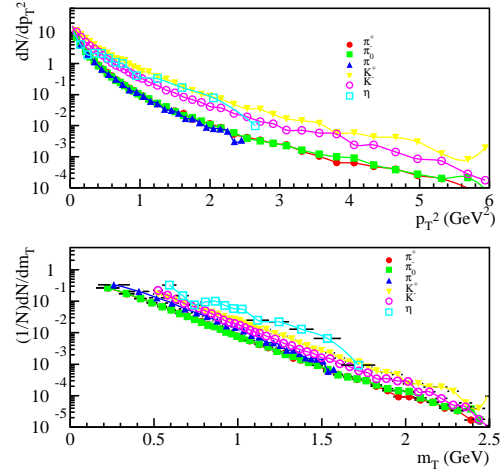}
\caption{Measured spectra of six identified particles in pp collisions at
$\sqrt s =27.4$ GeV \cite{na27}. The measured $\d n /\d p_T^2$ spectra (above) 
have been normalised so as to have the same value at $p_T =0$. While slopes 
of $\d n /\d p_T^2$ spectra are different, those of $\d n /\d m_T$ (below) 
are approximately the same for all identified particles.}
\label{mtslope}
\end{figure}

\section{Method of data analysis}

A peculiar prediction of a statistical picture in high energy collisions 
which is relevant to transverse momentum spectra is the so-called $m_T$ 
scaling: there should be an apparent common slope for $\d n/dm_T$ 
spectra of identified hadrons at a given centre-of-mass energy. By looking 
at eq.~(\ref{two}) one can easily realize that this holds in the limit
$\utm \rightarrow  0$. In principle, $m_T$ scaling applies only to primary 
hadrons, namely those directly emitted from the hadronising source. On the 
other hand, as stated in Sect.~1, most observed or reconstructed particles 
in experiments arise from decays of heavier hadrons. Those secondary decays 
may well distort the primary spectrum shape, thus spoling $m_T$ scaling.
Nevertheless, it can be seen in fig.~(\ref{mtslope}) that $m_T$ scaling 
apparently also holds for measured final hadrons, at least in the examined 
collision, implying seemingly little distortion from primary to final spectrum.
Settling this and other related issues demands a thorough analysis taking 
into account the effect of decays. Therefore, measured $p_T$ spectrum of 
the $j^{\rm th}$ hadron species should be compared with the sum of its primary 
spectrum and the contribution arising from all heavier hadrons decaying into it:  

\begin{equation}\label{three}
\frac{\d n_j}{\d p_T}  = \frac{\d n_j}{\d p_T}\vl^{\rm primary}
+ \sum_k \frac{\d n_j}{\d p_T}\vl^{k \rightarrow j}
\end{equation}
Whilst $\d n_j/\d p_T|^{\rm primary}$ is given by eq.~(\ref{two}), the calculation 
of the second term in the right hand side of eq.~(\ref{three}) within a 
statistical-thermal framework is a very complicated problem as the number of 
contributing resonances and decay modes is huge and also owing the presence of 
cascade decays (e.g. $\eta' \rightarrow \rho \rightarrow \pi$). 
In refs.~\cite{sollfran,sollfran2,wiedheinz} an analytical
approach has been taken including two and three body decays of directly 
emitted resonances. On the other hand, we have adopted \cite{provid} a 
mixed numerical-analytical method which allows a very accurate and exhaustive
calculation. Our method can be summarised as follows: for any couple of 
hadrons $(j,k)$, where $k$ decays into $j + X$, either directly or through 
a cascade process, we have determined the momentum spectrum 
$\d n_j/\d p |^{k \rightarrow j}_{\bb=0}$ in the equivalent cluster's rest
frame via a Monte-Carlo method by randomly generating 200,000 $k$'s decays, 
according to phase space, by picking $k$'s three-momenta from a thermal 
distribution with a given temperature $T$:

\begin{equation}\label{four}
\frac{\d n_k}{\d^3 p}\vl_{\bb=0} \propto \exp{\{-\sqrt{p^2+m_k^2}/T\}}
\end{equation}  
Then, we have boosted the obtained 
$\d n_j/\d p |^{k \rightarrow j}_{\bb=0}$ distributions in order to compute the 
$\d n_j/\d p |^{k \rightarrow j}(\utm)$ for a given transverse four-velocity 
$\utm$ of the equivalent cluster, to be plugged into eq.~(\ref{three}), with 
an integral formula derived by the authors \cite{prep}:

\begin{eqnarray}\label{five}
&& \!\!\!\!\!\!\!\!\!\!\! \frac{\d n_j}{\d p_T}\vl^{k \rightarrow j}
\!\!\!\!\!\!\!\!\!(\utm) =  
\frac{4 p_T}{\sqrt{1+\utm^2}\,m_T} \int_0^\infty \d p' 
\frac{\d n_j}{\d p'}\vl_{\bb=0}^{k \rightarrow j} \nonumber \\
&& \times \; \frac{1}{2 \pi p' \sqrt{(z_+ - z_{\rm min})(z_{\rm max}+1)}} 
\; {\rm F}\big( \frac{\pi}{2},r \big)
\end{eqnarray}
where F is the elliptic integral of the first kind and:

\begin{eqnarray}\label{six} 
 && r = \sqrt{\frac{(z_+ - z_{\rm max})(z_{\rm min}+1)}{(z_+ - z_{\rm min})(z_{\rm max}+1)}}
  \nonumber \\ 
 && z_{\rm max} = {\rm max}(1,z_-) \qquad z_{\rm min} = {\rm min}(1,z_-) 
  \nonumber \\ 
 && z_{\pm}= \frac{\epsilon' \pm \utm p_T}{\sqrt{1+\utm^2}\, m_T} \;\;
 \epsilon' = \sqrt{p'^2+m_j^2}
\end{eqnarray} 
\begin{table*}[ht]
\caption{Results of the fit to hadron average multiplicities for the chosen set of
hadronic and \ee collisions. The $\chi^2$ value in pp collisions is very high due
to lack of systematic errors \cite{chlia} in the measurements of particle multiplicities 
\cite{agui}.}
\label{table:1}
\newcommand{\m}{\hphantom{$-$}}
\newcommand{\cc}[1]{\multicolumn{1}{c}{#1}}
\renewcommand{\tabcolsep}{1pc} 
\renewcommand{\arraystretch}{1.2} 
\begin{tabular}{@{}llllll}
\hline
  collision&$\sqrt s\,$(GeV)& $T\,$ (MeV)  & $VT^3$        & \ssb          & $\chi^2/dof$ \\
\hline
 K$^+$p   & 11.5           &176.9$\pm$2.6 & 5.85$\pm$0.39 &0.347$\pm$0.020 & 68.0/14       \\ 
 $\pi^+$p & 21.7           &170.5$\pm$5.2 & 10.8$\pm$1.2  &0.734$\pm$0.049 & 39.7/7       \\
 K$^+$p   & 21.7           &175.8$\pm$5.6 & 8.48$\pm$1.05 &0.578$\pm$0.056 & 38.0/9       \\  
 pp       & 27.4           &162.6$\pm$1.6 & 14.17$\pm$0.66&0.644$\pm$0.018 & 313.9/29     \\ 
\hline
  collision&$\sqrt s\,$(GeV)& $T\,$ (MeV)  & $VT^3$        & $\gamma_S$    & $\chi^2/dof$ \\
\hline   
 \ee       & 14             &167.3$\pm$10.4& 9.7$\pm$2.6 &0.795$\pm$0.089 & 1.5/3       \\               
 \ee       & 22             &172.5$\pm$5.3 &10.6$\pm$1.8 &0.767$\pm$0.091 & 1.0/3       \\                 
 \ee       & 29             &159.0$\pm$1.8 &17.3$\pm$1.1 &0.710$\pm$0.039 & 29.3/12       \\                 
 \ee       & 43             &162.5$\pm$5.4 &16.2$\pm$2.1 &0.768$\pm$0.064 & 3.0/3       \\               
\hline
\end{tabular}\\[2pt]
\end{table*}
The main advantage of formula (\ref{five}) is to allow a quick computation of
transverse momentum spectra associated with particle decays for any tranverse 
four-velocity once the corresponding momentum spectra at zero velocity are 
known. To obtain it, we took advantage of the isotropy of 
$\d n_j/\d p |^{k \rightarrow j}_{\bb=0}$ distributions, which holds as 
long as decay products distribution is isotropic, which is true for 
simple phase-space decays with no polarisation. The integration in 
the variable $p'$ in eq.~(\ref{five}) has been performed numerically. 
The Monte-Carlo generation of $\d n_j/\d p |^{k \rightarrow j}_{\bb=0}$ 
spectra, depending on the temperature, has been performed for all unstable 
hadrons and resonances up to a mass of 1.8 GeV and it has been repeated for 
temperature values between 140 and 190 MeV in steps of 1 MeV, so as to have a 
large enough range of temperatures where to search for minima in a fitting 
procedure. For any other temperature value, the spectra have been calculated 
by means of a linear interpolation. Overall, for hadronic collisions, we 
have generated about 1200 spectra associated to 144 unstable hadrons for 
each $T$ value.\\        
The abundances of all hadrons, which are needed to assess the 
contribution of secondaries in the spectra of eq.~(\ref{three}), have been 
preliminarly set to the model values calculated by fitting the parameters 
$T$, $V$ and \ssb or $\gamma_S$ to the experimentally measured yields.
This procedure amounts to exclude the overall normalisation dependence of 
$\d n_j /\d p_T |^{k \rightarrow j}$ on $T$ in the fit so to 
keep only its shape dependence on $T$.\\
Whilst the $\gamma_S$ parameter has already extensively been used in 
multiplicity fits \cite{beca1,beca2,beca3}, a new parameter \ssb is 
introduced here for the strangeness suppression. The particle yields are 
computed by constraining
the number of newly created \ss pairs to fixed integers which fluctuate 
according to a Poissonian distribution. The mean value of the Poissonian
distribution, namely the mean value of newly created \ss pairs, is the
adjustable fit parameter \ssb. For the present, this new parametrisation 
has been used only for hadronic collisions (see Table 1) because \ee 
collisions present difficulties related to the production of heavy flavoured 
quarks which makes calculations extremely slow. The mathematical 
features of the \ssb parametrisation are described in detail in 
ref.~\cite{prep}.\\          
The fitted values, which update those in refs.~\cite{beca2,beca3} for 
hadronic and \ee collisions are quoted in Table 1 \footnote{The 
measurements used to perform this fit have been gathered through
the Durham reaction data database \cite{durh}; the relevant 
references will be quoted in ref.~\cite{prep}.}. The fitted temperature 
in pp collision in the present analysis is somewhat lower compared with 
ref.~\cite{beca3} owing to an updated set of hadron parameters \cite{pdg},
an upgraded procedure of $\chi^2$ minimisation also taking into account 
correlations in the effective variance method \cite{beca5}, as well 
as the replacement of $\gamma_S$ with \ssb. The fitted temperature values
show a very interesting trend consisting in a slight but significant increase
as centre-of-mass energy diminishes; this is in agreement with the observation
pointed out in ref.~\cite{goren}.
\begin{figure}[htb]
\vspace{5pt}
\includegraphics[width=17pc]{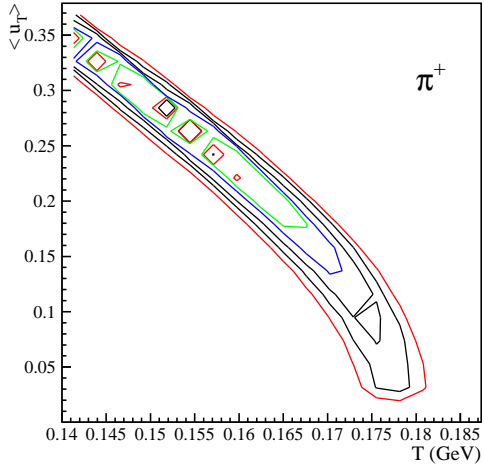}
\caption{$\chi^2$ contour plot in the $T-\utm$ plane for $\pi^+$ $p_T^2$
spectrum in pp collisions at $\sqrt s =$ 27.4 GeV.}
\label{picont}
\end{figure}

\section{Data set and results for hadronic collisions}

\begin{figure}[htb]
\vspace{5pt}
\includegraphics[width=17pc]{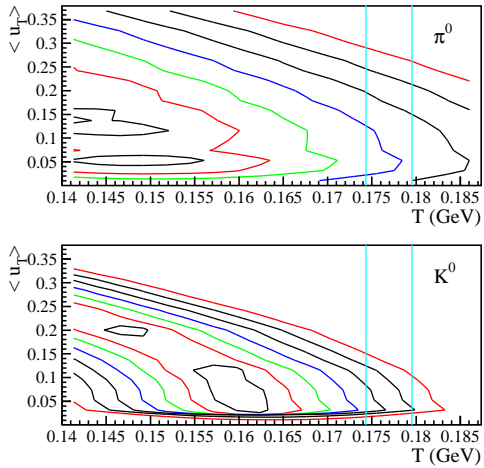}
\caption{$\chi^2$ contour plot in the $T-\utm$ plane for $\pi^0$ and K$^0_S$ 
$p_T^2$ spectra in K$^+$p collisions at $\sqrt s =$ 11.5 GeV. The two vertical 
lines define the 1-sigma band corresponding to the temperature fitted by using 
average particle multiplicities. Local minima lie outside the 1-sigma temperature 
band.}
\label{kpcontlow}
\end{figure}

The collisions and relevant centre-of-mass energies to be analysed have been 
primarily chosen on the basis of a fairly large number of measured particle 
yields and spectra. Particle multiplicities are needed to establish
with some accuracy the contribution of secondary decays while many identified
hadron spectra allow performing a test of universality on the slopes governed
by $T$ and $\utm$. This criterion led to singling out mainly four collision 
systems for hadronic collisions: K$^+$p at $\sqrt s = 11.5$ \cite{kp070}
and 21.7 GeV \cite{na22kp}, $\pi^+$p at $\sqrt s = 21.7$ GeV \cite{na22gp}
and pp $\sqrt s = 27.4$ GeV \cite{na27}. For each identified particle transverse
momentum spectrum, three free parameters have been determined by fitting 
eq.~(\ref{three}) to the data: $T$, $\utm$ and an overall normalisation parameter 
$A$. The fits have been performed in the variables $p_T$ or $p_T^2$ according to the 
available experimental spectrum; henceforth, we will denote by $\d n/ \d p_T$ 
both $\d n/ \d p_T$ and $\d n/ \d p_T^2$. As we have mentioned before, overall 
multiplicities of hadrons in eq.~(\ref{three}) have been fixed to model values 
by using fit parameters in Table 1.\\
The fit is performed by minimising the following $\chi^2$:
\begin{equation}\label{chi2}
  \chi^2 = \!\!\! \sum_{p_T \rm bins} \frac{\Big[\int_{{\rm bin}_i} \d p_T \, \d n/ \d p_T - 
  {\rm (Exp.value)}_i\Big]^2}{\sigma_i^2}  
\end{equation}
The errors $\sigma_i$ in the denominator of eq.~(\ref{chi2}) have been taken 
as the sum in quadrature of experimental errors and errors arising from the 
uncertainty on masses, widths and branching ratios of hadrons decaying into the 
examined hadron, according to the effective variance method (see ref.~\cite{prep} 
for a detailed description).
\begin{figure}[htb]
\vspace{5pt}
\includegraphics[width=17pc]{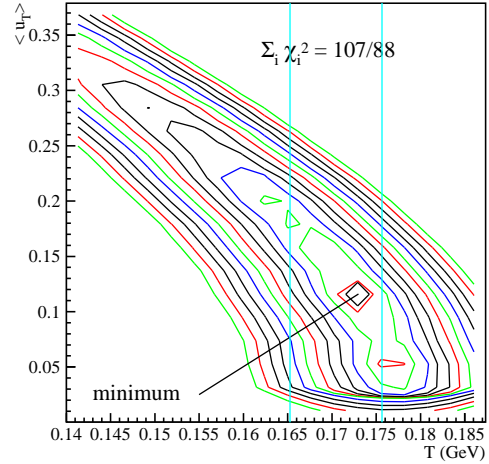}
\caption{Contour plot in the $T-\utm$ plane of $\sum_i \chi^2_i$ for $\pi^+$p
collisions $\sqrt s =$ 21.7 GeV. The identified particles involved in the sum 
are $\pi^-$, $\pi^0$, K$^0_S$, $\rho^0$, K$^{*0}$ and $\bar{\rm K}^{*0}$. The two
vertical lines define the 1-sigma band corresponding to the temperature fitted
by using average particle multiplicities. The $\sum_i \chi^2_i$ quoted in the 
picture is its value in the local minimum pointed by the solid line.}
\label{gpcont}
\end{figure}
\begin{figure}[htb]
\vspace{5pt}
\includegraphics[width=17pc]{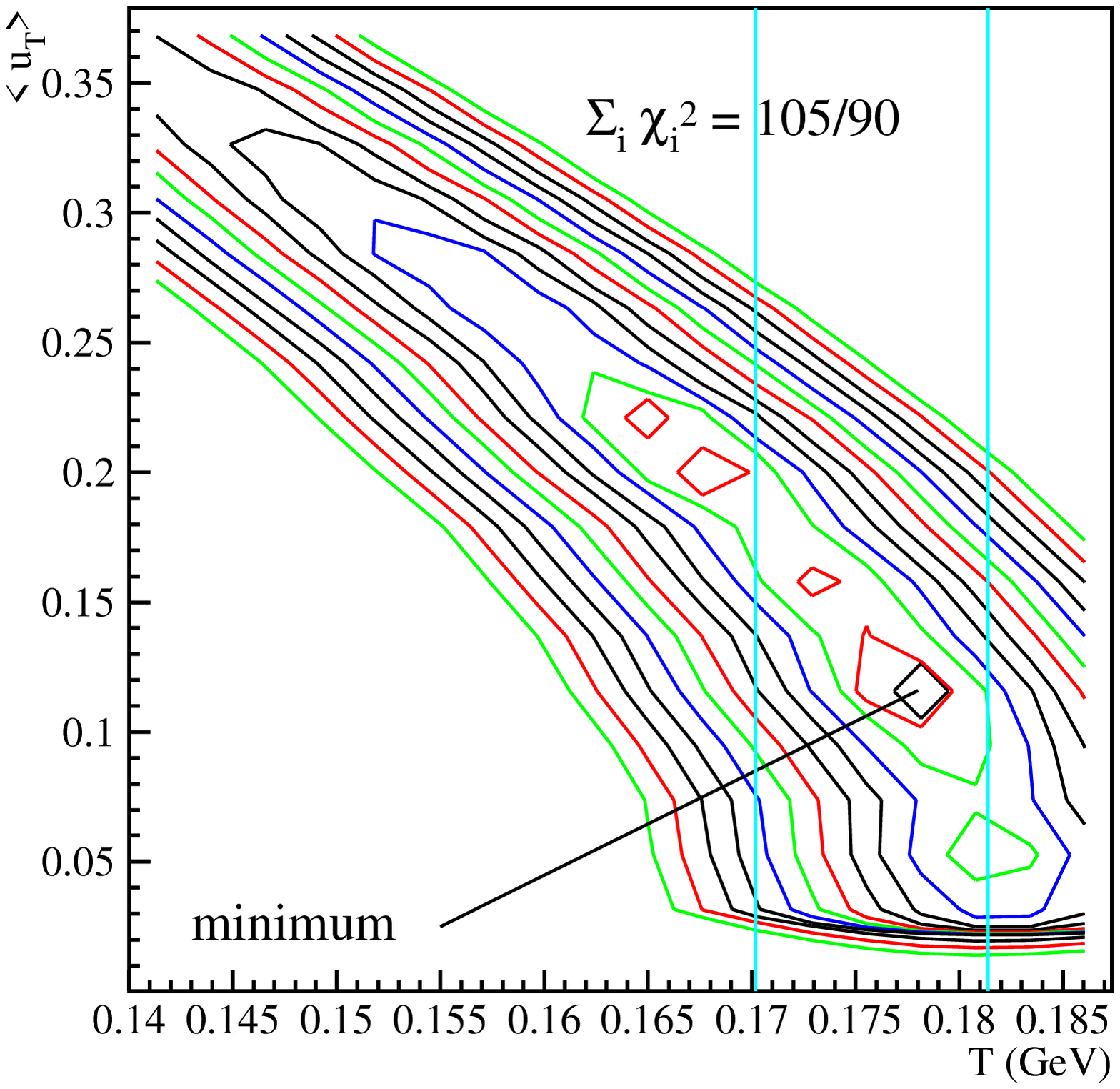}
\caption{Contour plot in the $T-\utm$ plane of $\sum_i \chi^2_i$ for K$^+$p
collisions $\sqrt s =$ 21.7 GeV. The identified particles involved in the sum 
are $\pi^-$, $\pi^0$, K$^0_S$, $\rho^0$, K${^*0}$ and f$_2(1270)$. The two
vertical lines define the 1-sigma band corresponding to the temperature fitted
by using average particle multiplicities. The $\sum_i \chi^2_i$ quoted in the 
picture is its value in the local minimum pointed by the solid line.}
\label{kpcont}
\end{figure}
\begin{figure}[htb]
\vspace{5pt}
\includegraphics[width=17pc]{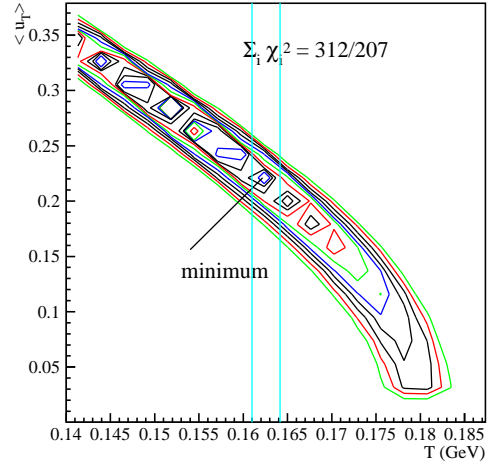}
\caption{Contour plot in the $T-\utm$ plane of $\sum_i \chi^2_i$ for pp
collisions $\sqrt s =$ 27.4 GeV. The identified particles involved in the sum 
are $\pi^+$, $\pi^-$, $\pi^0$, K$^+$, K$^-$, $\eta$, $\rho^0$, and f$_2(1270)$.
The two vertical lines define the 1-sigma band corresponding to the temperature 
fitted by using average particle multiplicities. The $\sum_i \chi^2_i$ quoted 
in the picture is its value in the local minimum pointed by the solid line. }
\label{ppcont}
\end{figure}
Once the normalisation parameter $A$ is fitted, a study of $\chi^2$ contour 
lines is performed in the $T-\utm$ plane in order to look for possible 
local minima. Indeed, the presence of several local minima along a band 
is a quite general feature of the $\chi^2$ (see fig.~(\ref{picont})) owing to 
the strong anticorrelation between $T$ and $\utm$. 
Hence, different solutions are possible for the $(T, \utm)$ pair. In order 
to enforce universality of such parameters among different hadron species we 
have chosen a solution for each particle according to the following procedure:
\begin{enumerate}
\item{} firstly, we have determined the contour lines of $\sum_{i=1}^N \chi^2_i$
as a function of $T$ and $\utm$ by fixing all normalisation parameters to
their fitted values, where $N$ is the number of different hadron species spectra
for a given collision. The minimisation of the sum of all $\chi^2$'s amounts
to fit a common value of $T$ and $\utm$ for all measured hadrons.
\item{} secondly, among the different local minima we have chosen the one lying 
in or closest to the band defined by $T$ fitted with average multiplicities 
along with its error (see figs.~(\ref{gpcont},\ref{kpcont},\ref{ppcont})). 
\item{} finally, all single particle spectrum fits have been repeated seeking
a local minimum as close as possible to the previously determined 
$\sum_{i=1}^N \chi^2_i$ minimum. 
\end{enumerate}
It must be noted that this procedure aims at achieving the best agreement between 
temperatures determined by using different observables, namely particle yields 
and transverse momentum spectra.
The $\sum_{i=1}^N \chi^2_i$ contour plots are shown in 
figs.~(\ref{gpcont},\ref{kpcont},\ref{ppcont}) for the examined collisions. 
There is a good agreement between the $T$ obtained from multiplicities
and the location of minima in the $T-\utm$ plane. For K$^+$p and $\pi^+$p at 
$\sqrt s = 21.7$ GeV the absolute minimum lies nearly about the centre of the band 
demonstrating an intriguing correlation between the slope of transverse momentum
distributions and the slope of average multiplicities as a function of mass.
Moreover, the $\utm$ values turn out to be small, thus justifying the series
expansion described in Sect.~2. Conversely, in K$^+$p collisions at 
$\sqrt s = 11.5$ GeV a clear disagreement emerges for 
K$^+$p at $\sqrt s = 11.5$ GeV between differently determined temperatures
(see fig.~(\ref{kpcontlow})) which can be explained with the inadequacy of our used 
parametrisation at low energy where exact $p_T$ conservation, neglected in
the canonical framework, is expected to play a significant role.
\begin{figure}[htb]
\vspace{5pt}
\includegraphics[width=17pc]{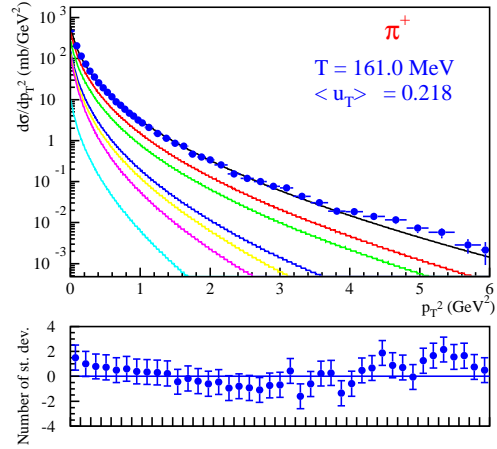}
\caption{Above: $\pi^+$ measured and fitted spectrum in pp collisions at 
$\sqrt s =$ 27.4 GeV. The lines lying below the top solid line represent
the cumulative contributions to the fitted spectrum of secondary $\pi^+$'s 
arising from the decays of doubly and singly strange baryons, other baryons, 
strange, charged and neutral mesons. Below: residuals distribution.}
\label{piplot}   
\end{figure}
\begin{figure}[htb]
\vspace{5pt}
\includegraphics[width=17pc]{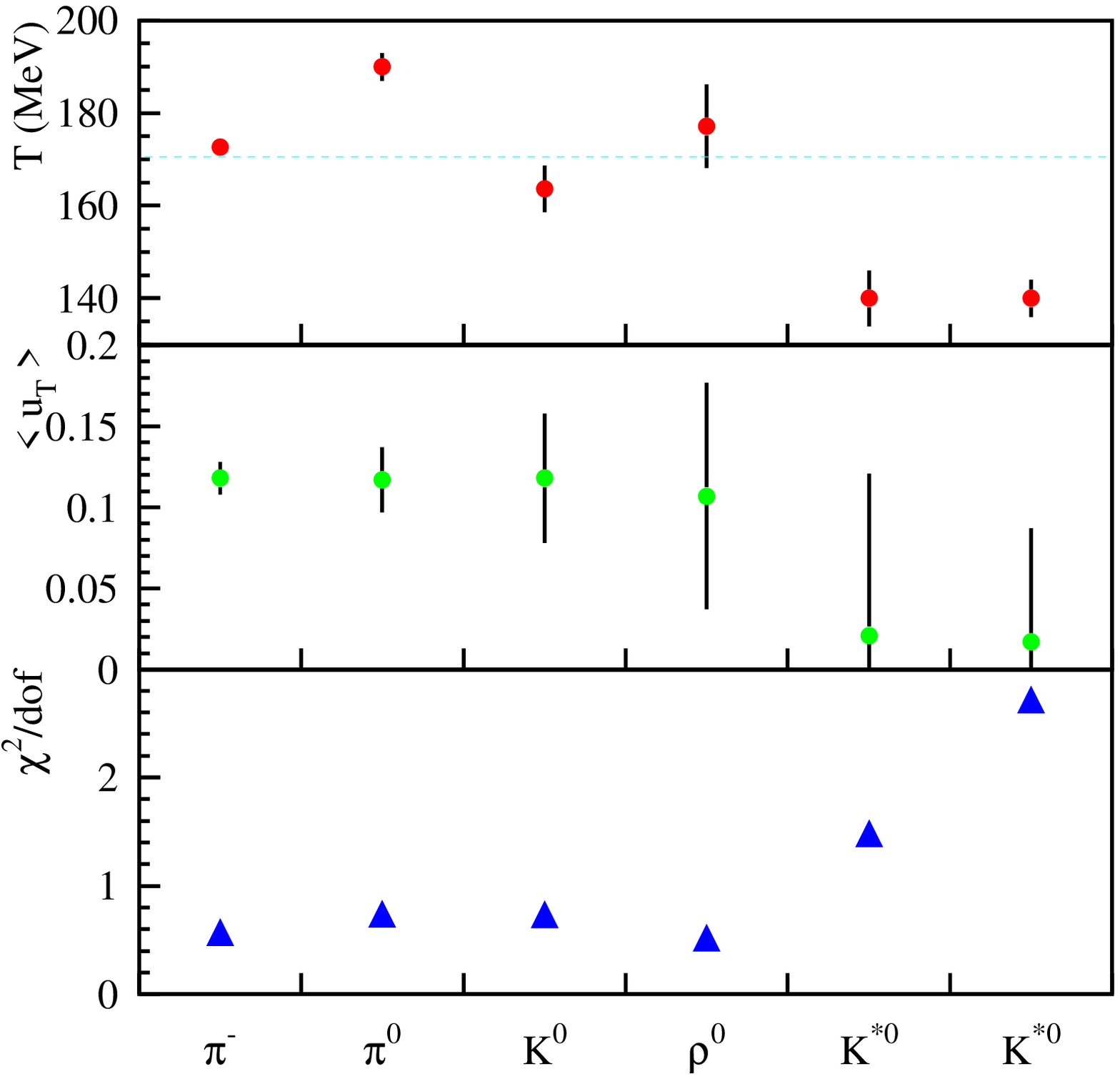}
\caption{Summary of the fits to transverse momentum spectra of identified
particles in $pi^+$p collisions at $\sqrt s =$ 21.7 GeV. From top to bottom: 
temperatures (the dashed line is the value obtained by fitting multiplicities), 
average transverse four-velocity $\utm$ and minimum 
$\chi^2/dof$. The errors on $T$ and $\utm$ are still very rough estimates.}
\label{summagp}
\end{figure}
\begin{figure}[htb]
\vspace{5pt}
\includegraphics[width=17pc]{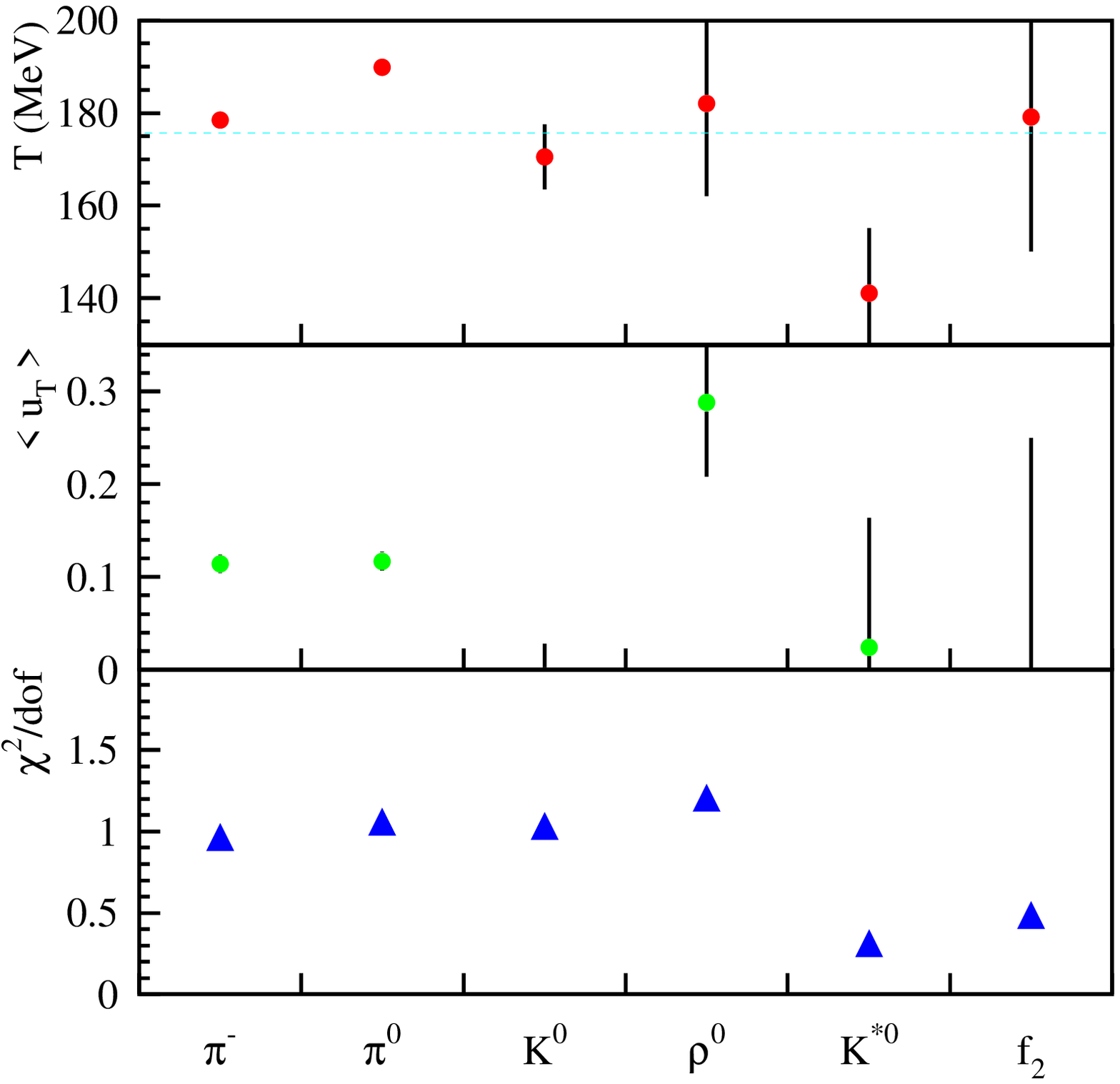}
\caption{Summary of the fits to transverse momentum spectra of identified
particles in K$^+$p collisions at $\sqrt s =$ 21.7 GeV. From top to bottom: 
temperatures (the dashed line is the value obtained by fitting multiplicities), 
average transverse four-velocity $\utm$ and minimum 
$\chi^2/dof$. The errors on $T$ and $\utm$ are still very rough estimates.}
\label{summakp}
\end{figure}
\begin{figure}[htb]
\vspace{5pt}
\includegraphics[width=17pc]{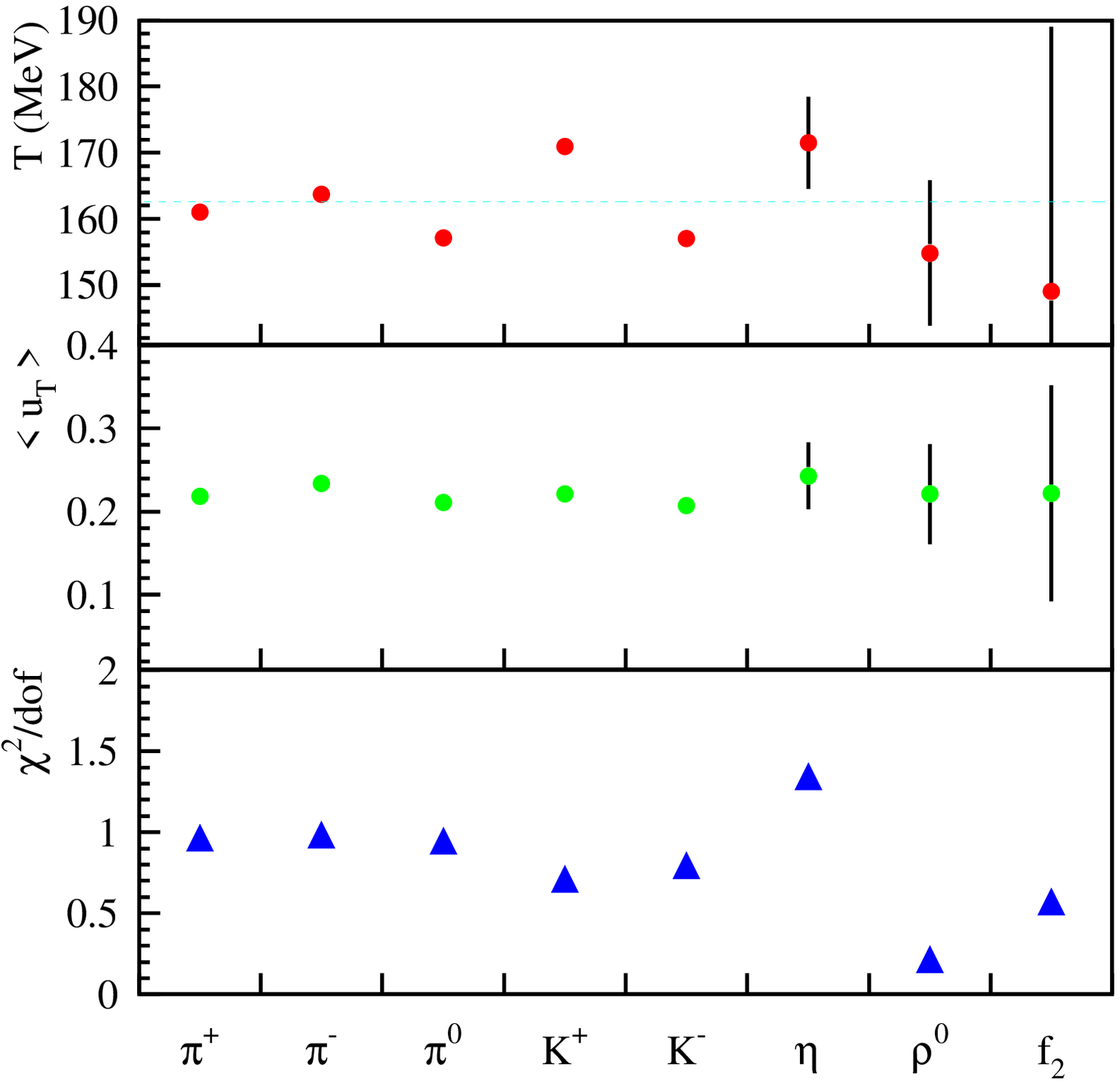}
\caption{Summary of the fits to transverse momentum spectra of identified
particles in pp collisions at $\sqrt s =$ 27.4 GeV. From top to bottom: 
temperatures (the dashed line is the value obtained by fitting multiplicities), 
average transverse four-velocity $\utm$ and minimum $\chi^2/dof$. The errors 
on $T$ and $\utm$ are still very rough estimates.}
\label{summapp}
\end{figure}
The summary plots of the single particle fits are shown in figs.~(\ref{summagp},
\ref{summakp},\ref{summapp}) while an example of fitted spectrum in pp collisions 
is shown in fig.~(\ref{piplot}). In general, very good fits (low $\chi^2$'s)
are obtained with good agreement between temperatures and average transverse
boost velocities fitted for different identified hadrons, especially in pp 
collisions. Some discrepancies show up in K$^+$p and $\pi^+$p collisions
where K$^*$'s seem to have slopes steeper than expected. It must be noted that
error estimates for $T$ and $\utm$ are still very rough.

\section{Results for \ee collisions}

We have also performed a similar analysis for \ee collisions at centre-of-mass
energies between 14 and 44 GeV. Compared with hadronic collisions, the analysis 
of \ee data presents several additional difficulties. First of all, the
production of heavy flavoured hadrons can not be neglected and the inclusion
of many more decay channels is implied. A major issue is the fact that $p_T$ 
is defined with respect to the so-called thrust or event axis. This direction 
is not known {\it a priori}, unlike in hadronic collisions
where $p_T$ is defined with respect to the beam line, and must be determined
on an event by event basis. Hence, whatever the algorithm being used, this
very fact introduces a bias on transverse momentum spectra because momenta 
projection are used to determine the event or thrust axis itself; often, this 
is done by just minimising the sum of particle $p_T$'s.
\begin{figure}[htb]
\vspace{5pt}
\includegraphics[width=17pc]{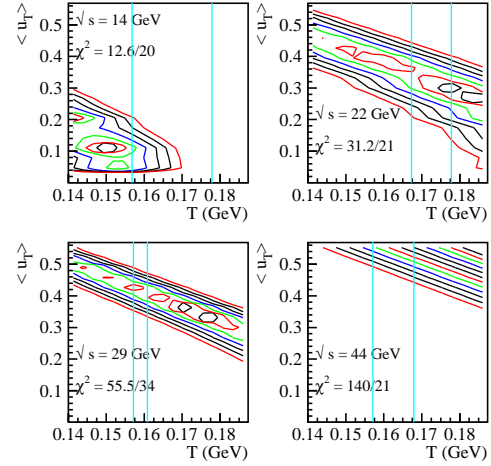}
\caption{$\chi^2$ contour plot in the $T-\utm$ plane for charged tracks 
$p_T$ or $p_T^2$ spectra in \ee collisions at four different centre-of-mass
energies ranging from 14 to 44 GeV. The two vertical lines define the 1-sigma 
band corresponding to the temperature fitted by using average particle 
multiplicities. For \ee collisions at $\sqrt s =$ 44 GeV, local minima lie
outside the considered $\utm$ upper bound and the minimum $\chi^2$ is poor.}
\label{eecont}
\end{figure}
In spite of this problem, we have repeated the same analysis performed
for hadronic collisions by using measured charged tracks transverse momentum
spectra \cite{eept} with respect to thrust axis. The $\chi^2$ contour plots 
are shown in fig.~(\ref{eecont}) along with $T$ bands obtained from multiplicity 
fits and resulting
$\chi^2$'s minima. It can be seen that no minimum is found within the band
for \ee collisions at $\sqrt s = 14$ GeV, a situation looking very similar
that found in K$^+$p collision at $\sqrt s = 11.5$ GeV. It is also found 
that fitted $\utm$ increases quite rapidly with centre-of-mass energy 
reflecting the rise of average $p_T$ of gluon radiation. Since our 
parametrisation of transverse momentum spectra with an average 
transverse four-velocity 
$\utm$ requires $\utm$ to be $\ll 1$ (see discussion in Sect. 2), it is
quite natural that fit quality deteriorates as centre-of-mass
energy increases. This is indeed found already at 44 GeV where $\utm > 0.55$
cannot longer be considered small. As energy increases, particle $p_T$
spectra become sensitive to the shape of clusters $u_T$ distribution 
(related in turn to radiated gluon $p_T$ spectrum) and not only to 
its mean value. Further on, they become dominated by the shape of gluon 
radiation spectrum whilst hadronisation $p_T$ plays the role of a small 
superimposed noise.
        
\section{Discussion and conclusions}

The analysis of transverse momentum spectra of several identified hadrons,
accurately taking into account the effect of hadron decays, indicates 
that they can be well reproduced within the statistical hadronisation
model for several high energy hadronic collisions. Furthermore, a good 
agreement is found between the temperatures
estimated by fits to average particles multiplicities and those extracted
by fitting transverse momentum spectra. This can be seen especially by
comparing figs.~(\ref{gpcont},\ref{kpcont},\ref{ppcont}); in fact, the 
location of the best $T$ value for transverse momentum spectra, determined 
by $\chi^2$'s lowest contour line, is apparently correlated to the best
$T$ value defined by particle multiplicities as spotted by the vertical
band. Indeed, those temperatures move toghether from $\simeq 162$ MeV 
in pp collisions at $\sqrt s =$ 27.4 GeV to $\simeq 175$ MeV in $\pi^+$p
and K$^+$p collisions at $\sqrt s =$ 21.7 GeV. In our view, this finding    
is a strong indication in favour of one of the key predictions of the 
statistical hadronisation 
model, namely the existence of a close relationship between the laws
governing the production of particles as a function of their mass and,
for each particle species, the production as a function of momentum 
(measured in the rest frame of the cluster they belong to) at the 
hadronisation.\\  
Besides this main conclusion, there are many remarks to be made. First
of all, our calculations assumed a canonical framework and so they are
expected to have only a limited validity at low centre-of-mass energies 
where the effect of exact transverse momentum conservation must show up.
Indeed, in \ee collisions at a centre-of-mass energy of 14 GeV and K$^+$p 
collisions at 11.5 GeV, a clear disagreement is found between the 
temperatures determined in the two fashions (see fig.~(\ref{kpcontlow})). 
Secondly, whilst the results obtained in pp collisions for different 
particles are in very good agreement among them, 
there are some significant discrepancies between different particles in 
$\pi^+$p and K$^+$p collisions, especially for K$^*$'s, which are not 
understood at the present. Finally, it is evident in the analysis of 
moderately high energy \ee collisions, at $\sqrt s =$ 44 GeV, that our 
parametrisation of transverse momentum spectra as a function of an average 
transverse four-velocity alone, is not accurate enough. As the average 
$p_T$ of radiated gluons and, as a consequence, of hadronising clusters, 
rises, particle spectra are influenced more by the shape of clusters 
transverse four-velocity spectrum than by primordial hadronisation $p_T$ spectrum 
so that more accurate calculations are needed involving perturbative QCD
to perform such an analysis. 
Finally, in the very high energy regime, $p_T$ spectra become insensitive 
to hadronisation and, as a consequence, it is no longer interesting to
study its properties. In summary, the presently used 
parametrisation of $p_T$ spectra is found to work well only in a limited 
centre-of-mass energy range (roughly between 20 and 30 GeV). At lower
energies, the study of $p_T$ is still of great interest in probing the
statistical features of hadronisation but complex microcanonical 
calculations are required. At higher energy, hadronisation is swamped 
by hard QCD dynamics in $p_T$ spectra and all it can do is to add 
a little smearing. 
  
\section*{Acknowledgments}

We would like to thank the organizers of the "Correlations and fluctuations"
conference in Torino. We are grateful to U. Heinz and U. A. Wiedemann for
clarifying discussions. Many thanks to J. A. Baldry for the careful revision
of the manuscript. This work would have not been possible without
the great Durham reaction data database; we wish to express our gratitude
to their editors and curators.



\end{document}